# Characterization of spin wave propagation in (111) YIG thin films with large anisotropy

A. Krysztofik,[1,b] H. Głowiński,[1,a] P. Kuświk,[1,2] S. Ziętek,[3] L. E. Coy,[4] J. N. Rychły,[5] S. Jurga,[4] T. W. Stobiecki,[3] J. Dubowik[1]

[1]Institute of Molecular Physics, Polish Academy of Sciences, M. Smoluchowskiego 17, PL-60-179 Poznań, Poland

[2]Centre of Advanced Technology, Adam Mickiewicz University, Umultowska 89c, PL-61-614 Poznań, Poland

[3]Department of Electronics, AGH University of Science and Technology, Al. Mickiewicza 40, PL-30-059 Kraków, Poland

[4]NanoBioMedical Centre, Adam Mickiewicz University, Umultowska 85, PL-61-614 Poznań, Poland

[5]Faculty of Physics, Adam Mickiewicz University, Umultowska 85, PL-61-614 Poznań, Poland

[6]Faculty of Physics and Applied Computer Science, AGH University of Science and Technology, Al. Mickiewicza 30, PL-30-059 Kraków, Poland

[a]E-mail: hubert.glowinski@ifmpan.poznan.pl
[b]E-mail: adam.krysztofik@ifmpan.poznan.pl

## Abstract

We report on long-range spin wave (SW) propagation in nanometer-thick Yttrium Iron Garnet (YIG) film with an ultralow Gilbert damping. The knowledge of a wavenumber value $|\vec{k}|$ is essential for designing SW devices. Although determining the wavenumber $|\vec{k}|$ in experiments like Brillouin light scattering spectroscopy is straightforward, quantifying the wavenumber in all-electrical experiments has not been widely commented so far. We analyze magnetostatic spin wave (SW) propagation in YIG films in order to determine SW wavenumber $|\vec{k}|$ excited by the coplanar waveguide. We show that it is crucial to consider influence of magnetic anisotropy fields present in YIG thin films for precise determination of SW wavenumber. With the proposed methods we find that experimentally derived values of $|\vec{k}|$ are in perfect agreement with that obtained from electromagnetic simulation only if anisotropy fields are included.

Spin wave (SW) propagation in magnetic thin film structures has become intensively investigated topic in recent years due to promising applications in modern electronics [ 1, 2, 3, 4 ]. The wavenumber (or equivalently – the wavelength $\lambda = 2\pi/|\vec{k}|$) is an important parameter to account for propagation characteristics. For example, it is essential to choose SW wavenumber and correlate it to certain device dimension in order to ensure observation of expected phenomena in SW devices e.g. in magnonic crystals [ 5, 6 ] or devices based on wave interference such as SW transistor [ 2 ], SW logic gates [ 2 ], Mach-Zender type interferometers [ 7 ]. The knowledge of SW wavenumber is also very important in the assessment of the effective magnitude of Dzaloshinskii-Moriya interaction using collective spin-wave dynamics [ 8 ].

In propagating SW spectroscopy experiments two shorted coplanar waveguides (CPWs) are commonly used as a transmitter and a receiver [ 9 ]. Each CPW, integrated within the film, consists of a signal line and two ground lines connected at one end. When a rf-current flows through the transmitter it induces an oscillating magnetic field around the lines that exerts a torque and causes spin precession in the magnetic material beneath. The inverse effect is then used for SW detection by the receiver. Since the generated magnetic field is not homogenous with reference to the film plane and solely depends on CPW geometry, it determines the distribution of SW wavenumber that can be excited.

It is assumed that the transmitter excites a broad spectrum of SW wavevectors of wavenumber $k$ extending to $k_{max} \approx \pi/W$ ($W$ is a width of CPW line) with a maximum of excitation amplitude approximately around $k_{maxAmp} \approx \pi/2W$ [ 10 ]. The question now is: what is the actual wavenumber of the SW with the largest amplitude detected by the receiver situated at a certain distance from the transmitter. It appears that while in Brillouin light scattering spectroscopy $k$ is easily accessible, in all electrical spin wave spectroscopic experiments the determination of SW wavenumber is rather challenging [ 11 ].

We aim to answer this question by analyzing our experimental results of SW propagation in yttrium iron garnet ($Y_3Fe_5O_{12}$, YIG) thin films. YIG films are known as possessing the lowest Gilbert damping parameter enabling the SW transmission over the distances of several hundred micrometers [ 2, 12 ]. However, YIG films synthesized by pulsed laser deposition (PLD) exhibit substantially disparate values of anisotropy fields and saturation magnetization, depending on the growth process parameters and, consequently, stoichiometry of the obtained film [ 13, 14, 15 ]. It has already been theoretically predicted that anisotropy may significantly affect SW propagation and the transmission characteristics [ 16, 17 ]. Therefore, for such YIG films, SW spectra analysis requires careful consideration of anisotropic properties of a given film.

Here, we compare two methods of experimental determination of the SW wavenumber which include anisotropy fields. The experimental results are then compared with electromagnetic simulations.

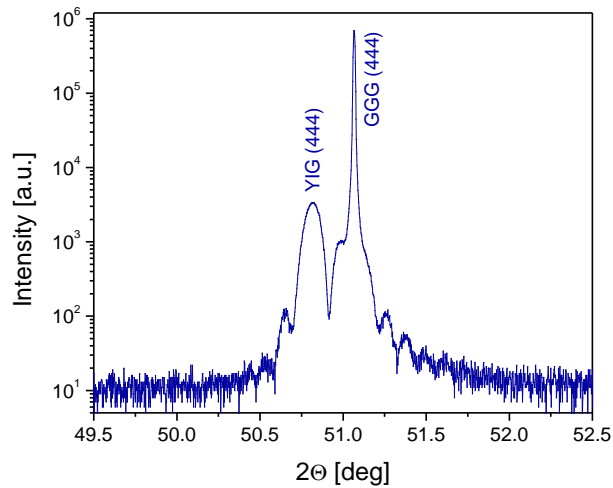

Fig. 1. A θ-2θ XRD scan of epitaxial YIG film on GGG (111) substrate near the GGG (444) reflection.

YIG film was grown on a monocrystalline, *111*-oriented Gadolinium Gallium Garnet substrate ($Gd_3Ga_5O_{12}$, GGG) by means of PLD technique. Substrate temperature was set to 650°C and under the $1.2 \times 10^{-4}\ mbar$ oxygen pressure ($8 \times 10^{-8}\ mbar$ base pressure) thin film was deposited at the $0.8\ nm/min$ growth rate using third harmonic of Nd:YAG Laser ($\lambda = 355\ nm$). After the growth, the sample was additionally annealed *ex situ* at 800°C for $5\ min$. X-ray diffraction and reflection measurements showed that the YIG film was single-phase, epitaxial with the GGG substrate with the thickness of $82\ nm$ and RMS roughness of $0.8\ nm$. XRD θ-2θ scan, presented in Fig. 1, clearly shows the high crystallinity of the YIG film, displaying well defined Laue oscillations, typical for highly epitaxial films, which clearly point to the high quality and well textured YIG (111) film [ 18 ]. Subsequently, a system of two CPWs made of $100\ nm$ thick aluminum was integrated onto YIG film (Fig.2) using a maskless photolithography technique. The width $W$ of signal and ground lines was equal to $9.8\ \mu m$ and the gaps between them were $4\ \mu m$ wide. The distance between the centers of signal lines was $150\ \mu m$.

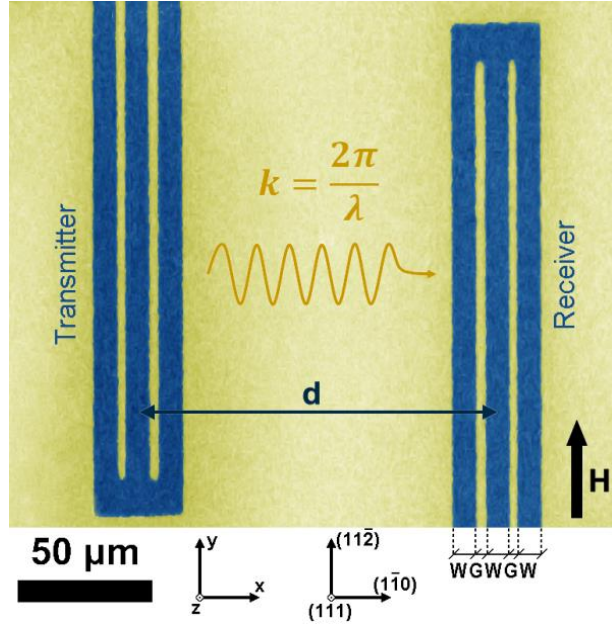

Fig. 2. SEM image of the integrated CPWs on the YIG film. The distance $d$ between the transmitter and the receiver is equal to 150 $\mu m$. The depicted Cartesian and crystallographic coordinate system is used throughout this paper. The width of signal and ground lines is marked with $W$. $G$ denotes the gap width between the lines.

To investigate SW propagation we followed approach presented in Ref. [ 9 ] and [ 12 ]. Using a Vector Network Analyzer transmission signal $S_{21}$ was measured for Damon-Eshbach surface modes with wavevector $\vec{k}$ perpendicular to the magnetization for magnetic fields ranging from $-310\ Oe$ to $+310\ Oe$ (Fig. 3(a)). Exemplary $S_{21}$ signals (imaginary part), which are shown in Figs 3(b) and (c), reveal a series of oscillations as a function of frequency with a Gaussian-like envelope corresponding to the excited SW wavenumber distribution. Figure 3(c) shows that frequency separation $\Delta f$ between two oscillation maxima differs noticeably in value depending on the magnetic field. The decrease in signal amplitude is also observed since SW decay length is inversely proportional to the frequency, so that the low-frequency SWs propagate further away [ 12, 19 ].

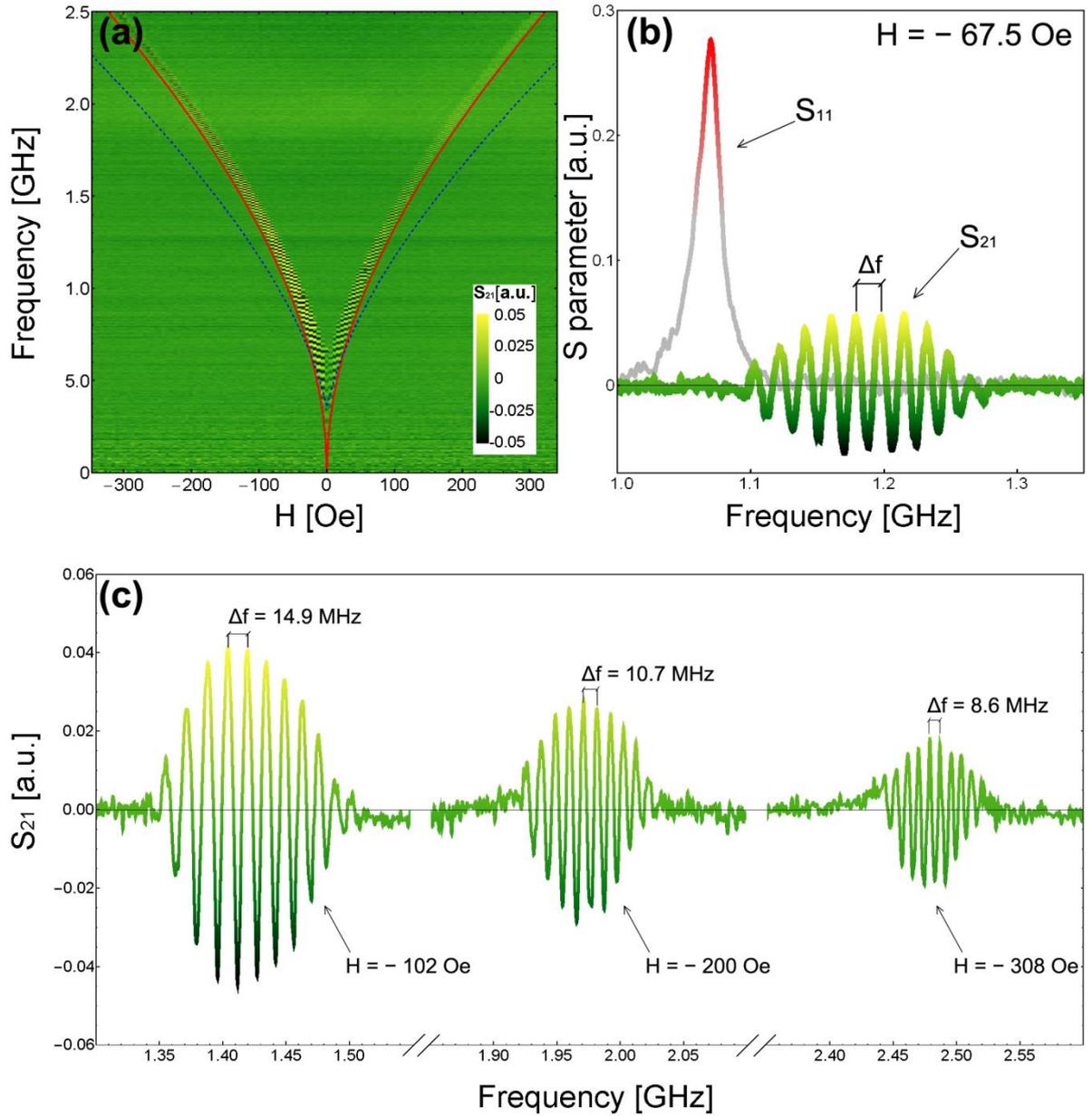

Fig. 3. (a) Color-coded SW propagation data $S_{21}$ measured at different magnetic fields. With a red line $f(H)$ dependence of the uniform excitation ($k = 0$) is depicted. The red line corresponds to the maximum in $S_{11}$ signal in (b). The blue dashed line represents a dispersion relation with $H_a = H_u = 0$. (b) Reflection ($S_{11}$, $k = 0$) and transmission ($S_{21}$, $k \neq 0$) signals. The plot illustrates a magnified cross-section of (a) at $H = -67.5\ Oe$. (c) SW spectra measured at different magnetic fields. Color-coding in (b) and (c) corresponds to the one defined in (a).

For the frequencies of the highest signal amplitude, the wavenumber $k_{maxAmp}$ can be determined according to the dispersion relation derived for (111) crystalline orientation of the YIG film [ 16, 17 ]:

$$f = \frac{\mu_B}{2\pi\hbar} g \sqrt{(H + 2\pi M_s tk)\left(H - \frac{1}{2}H_a - H_u + 4\pi M_s - 2\pi M_s tk\right) - \frac{1}{2}(H_a \sin(3\phi))^2}, \quad (1)$$

where $f$ is the microwave frequency, $\mu_B$ – the Bohr magneton constant, $\hbar$ – the reduced Planck constant, $g$ – the spectroscopic splitting factor, $H$ – the external magnetic field, $M_s$ – the saturation magnetization, $t$ – the film thickness, $k$ – the wavenumber, $H_a$ – the cubic anisotropy field and $H_u$ – the out-of-plane uniaxial anisotropy field. $H_a = \frac{2K_a}{M_s}$ and $H_u = \frac{2K_u}{M_s}$, where $K_a$ and $K_u$ are anisotropy constants. It should be highlighted that when $H_a = H_u = 0$, Eq. 1 becomes equivalent to the one originally obtained by Damon and Eshbach [20]. The azimuthal angle $\phi$ defines the in-plane orientation of magnetization direction with respect to the $(11\bar{2})$ axis of YIG film. In our study the term $-\frac{1}{2}(H_a \sin(3\phi))^2$ in Eq. 1 vanishes since magnetic field $H$ is parallel to $(11\bar{2})$ axis and $\phi = 0°$.

As can be seen from Eq. 1, in order to determine wavenumber $k$ one needs to evaluate many material constants, namely $g, M_s, t, H_a, H_u$ in the first instance. This problem can be partially solved with a broadband ferromagnetic resonance measurement of the film. For $k = 0$ Eq.1 simplifies to the formula, which allows for the determination of the spectroscopic factor $g$ and the effective magnetization $4\pi M_{eff}^* = -\frac{1}{2}H_a - H_u + 4\pi M_s$:

$$f_{k=0} = \frac{\mu_B}{2\pi\hbar} g \sqrt{H(H + 4\pi M_{eff}^*)}. \quad (2)$$

Therefore, within this approach, the film thickness and the saturation magnetization should be determined using other experimental methods.

To investigate ferromagnetic resonance of the YIG film, the reflection signal $S_{11}$ was measured. In order to avoid extrinsic contribution to the resonance linewidth caused by non-monochromatic excitation of the CPW ($2\pi\Delta f_{extr} = v_g \Delta k$) [21] and, consequently, possible ambiguities in the interpretation of resonance peak position, it is recommended to perform this measurement with the use of a wide CPW. Note that the full width at half maximum of a CPW excitation spectra $\Delta k \approx k_{maxAmp}$ [21]. In our study we used a CPW with signal and ground lines of the width equal to $450\ \mu m$ and with the $20\ \mu m$ wide gaps between them. For such a CPW, the simulated value of $k_{maxAmp}$ is equal to $49\ cm^{-1}$ and, therefore, yields negligible broadening that is of the order of a few MHz.

The measured $S_{11}$ signal (imaginary part) is depicted in Fig. 3(a) with the red line. It appears to lie just below the $S_{21}$ signal. Fitting to the experimental data with Eq. 2 gave following value of the spectroscopic factor $g = 2.010 \pm 0.001$ and the effective magnetization $M^*_{eff} = 169 \pm 7\ emu/cm^3$. A comparison of $4\pi M^*_{eff}$ with $4\pi M_s$ ($M_s = 120 \pm 19\ emu/cm^3$ was measured using Vibrating Sample Magnetometry) gives $-\frac{1}{2}H_a - H_u$ of $616\ Oe$, showing the substantial difference between obtained values of $M^*_{eff}$ and $M_s$. The determined value of $-\frac{1}{2}H_a - H_u$ remains in the midst of the range reported for PLD-grown YIG thin films, from $229\ Oe$ up to $999\ Oe$ [ 14, 22 ]. It is worth to mention that for fully stoichiometric, micrometer-thick YIG films made by means of liquid phase epitaxy (LPE) technique $-\frac{1}{2}H_a - H_u = 101\ Oe$ [ 14 ]. From the analysis of resonance linewidth vs. frequency [ 23 ] we additionally extracted Gilbert damping parameter of the YIG film, which equals to $\alpha = (5.5 \pm 0.6) \times 10^{-4}$ and implies low damping of magnetization precession.

Substitution of the $g$, $M^*_{eff}$, $M_s$ and $t$ values into Eq. 1 enabled the determination of wavenumber $k_{maxAmp} = 1980 \pm 102\ cm^{-1}$. It should be noted that if anisotropy fields were neglected in the Eq.1 ($H_a = H_u = 0$), yet only saturation magnetization was taken into account, a fitting to the experimental data would not converge. The calculated dispersion relation with the derived value of $k_{maxAmp}$, assuming $H_a = H_u = 0$ is depicted with blue dashed line in Fig. 3 (a). Omission of anisotropy fields in magnetization dynamic measurements may therefore lead to the significant misinterpretation of experimental results for YIG thin films.

Typical values of cubic magnetocrystalline anisotropy field $H_a$ range from $-18\ Oe$ to $-64\ Oe$ for PLD grown YIG films [ 14, 15, 22 ], what indicates that resonance measurements as well as spin wave propagation are governed by the out-of-plane uniaxial anisotropy. For the film employed in our study, the $H_u$ value is of about $-600\ Oe$ in agreement with previous reports [ 14, 15, 22 ]. For any more complex architecture of magnonic waveguides and circuits it is likewise imperative to investigate the in-plane anisotropy properties [ 24 ]. As can be seen from Eq. 1 one would expect a six-fold anisotropy in the plane of (111)-oriented single crystals, that is common among rare-earth substituted YIG garnets and LPE-YIG films [ 18, 25, 26, 27 ]. To examine this issue, we performed VSM and angular resolved ferromagnetic resonance measurements. Hysteresis loops for all measured in-plane directions exhibit no substantial differences regarding coercive field ($\approx 1.2\ Oe$), saturation field and saturation magnetization (Fig. 4(a)). The angular resolved resonance measurements confirm this result and show that the (111) YIG film is isotropic in the film plane (Fig. 4(b)). The main reason for this behavior is the low value of cubic anisotropy field which causes the resonance frequency modulation by a value of the fraction of MHz. Such small differences do not surpass the experimental error, nor would they significantly affect the coherent SW propagation. It is expected that the SW propagation characteristics, measured for any other crystallographic orientation, would therefore remain unaltered.

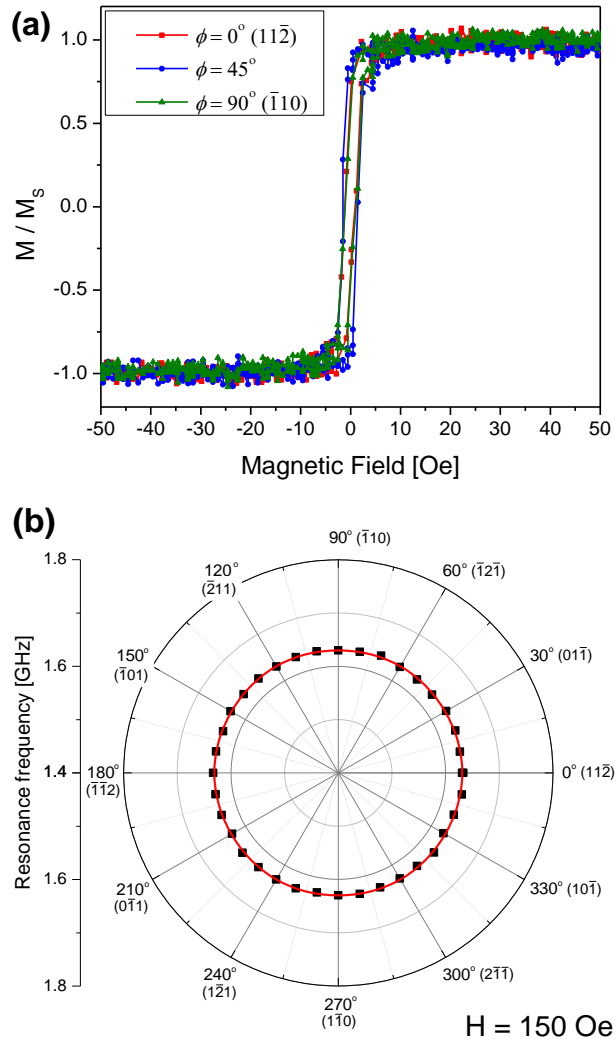

Fig. 4. (a) VSM hysteresis loops measured in the film plane for three different crystallographic directions. The magnetization is normalized to the saturation magnetization $M_s = 120 \pm 19\ emu/cm^3$. A paramagnetic contribution of the GGG substrate was subtracted for each loop. (b) Resonance frequency as a function of azimuthal angle $\phi$ taken at $H = 150\ Oe$. The red line depicts the calculated values of resonance frequency according to Eq.1 for $k = 0$, $H_a = -30\ Oe$ and $H_u = -600\ Oe$.

Another method of extracting SW wavenumber involves the analysis of the SW group velocity $v_g$. Following Ref. [ 21 ], $v_g$ can be determined from frequency difference $\Delta f$ between two oscillation maxima in S$_{21}$ signal according to the relation:

$$v_g = d\Delta f, \quad (3)$$

where $d$ is the distance between two CPWs. To determine $\Delta f$ we chose two neighboring oscillation maxima of the highest S$_{21}$ signal amplitude as it is shown in Fig. 3(b) and (c).

In Fig. 5 the derived values of group velocity are shown as a function of magnetic field. It is found that $v_g$ reaches the value of $7.6\ km/s$ for the field of $1.3\ Oe$ (preferable in magnonic information processing devices of high efficiency) and $1.4\ km/s$ for the field of $285\ Oe$. It should be highlighted that such big differences in $v_g$ values can be further utilized to design tunable, impulse-response delay lines as $v_g$ changes up to five times with the magnetic field. At a distance of $150\ \mu m$ between CPWs it would allow to achieve 20 to 110 $ns$ delay times of an impulse.

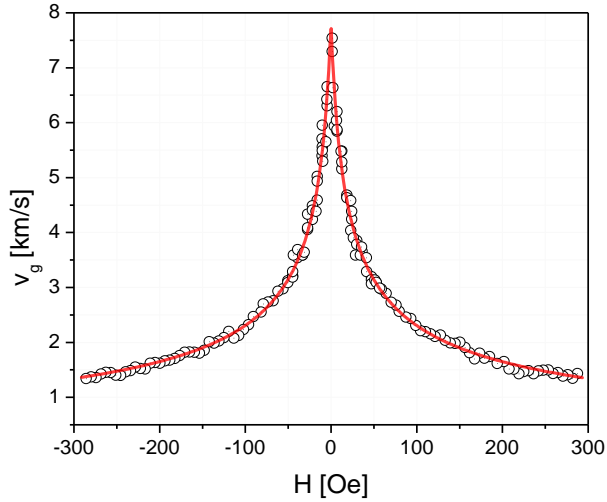

Fig. 5. Spin wave group velocity as a function of the external magnetic field. The red line represents a fit according to Eq. 4.

With the red line in Fig. 5 a fitting is depicted according to:

$$v_g = 2\pi\frac{\partial f}{\partial k} = \frac{\mu_B}{\hbar} g \frac{2\pi M_s t\left(-\frac{1}{2}H_a - H_u + 4\pi M_s - 4\pi M_s tk\right)}{2\sqrt{(H + 2\pi M_s tk)\left(H - \frac{1}{2}H_a - H_u + 4\pi M_s - 2\pi M_s tk\right)}}. \quad (4)$$

The main advantage of extracting SW wavenumber from $v_g(H)$ dependence is that it does not require additional measurement of $M_s$ which is often notably influenced by an error in the estimated film volume. Since the saturation magnetization $M_s$ can be treated as a fitting parameter in Eq. 4, the derivation of SW wavenumber involves only S$_{11}$, S$_{21}$ and thickness measurements. The determined values of $k_{maxAmp} = 1690 \pm 53\ cm^{-1}$ and $M_s = 116 \pm 2\ emu/cm^3$ remain in a good agreement with that obtained above - directly derived from dispersion relation ($k_{maxAmp} = 1980 \pm 102\ cm^{-1}$, $M_s = 120 \pm 19\ emu/cm^3$).

As can be seen from Figure 5, SW group velocity attains the maximum value as the magnetic field approaches $H = 0$. The maximum value of $v_g$ is given by:

$$v_g^{(H=0)} \cong \frac{\mu_B}{\hbar} g \sqrt{\frac{\pi M_s t}{2k} \left(-\frac{1}{2}H_a - H_u + 4\pi M_s[1 - tk]\right)}. \tag{5}$$

The zero-field region may therefore become the subject of interest for magnonic applications. Moreover, Eq. 5 shows that the maximum value of $v_g$ depends on the anisotropy fields. PLD-grown YIG films possessing a high anisotropy would allow faster information processing in SW circuits than LPE films for which the value of $-\frac{1}{2}H_a - H_u$ is smaller (as it was pointed out above).

To confront our experimental results with the expected, theoretical value of $k_{maxAmp}$, we performed electromagnetic simulations in *Comsol Multiphysics*. Here, CPW was modeled according to the geometry of the performed CPW (Fig. 2), assuming lossless conductor metallization, relative permittivity of the substrate $\varepsilon_r = 12$ and $50\ \Omega$ port impedance. From the simulated in-plane distribution of the dynamic magnetic field $h_x$ (inset of Fig. 6) an excitation spectra of CPW was obtained using discrete Fourier transformation of $h_x(x)$. The highest excitation strength is observed for $k_{maxAmp} = 1838\ cm^{-1}$, which corresponds well to the experimentally obtained values within 7% accuracy. The second observed maxima is at $k_2 = 6770\ cm^{-1}$. However, as its amplitude is 20 times lower with respect to the amplitude of $k_{maxAmp}$ it is not observed in the measured S$_{21}$ signal.

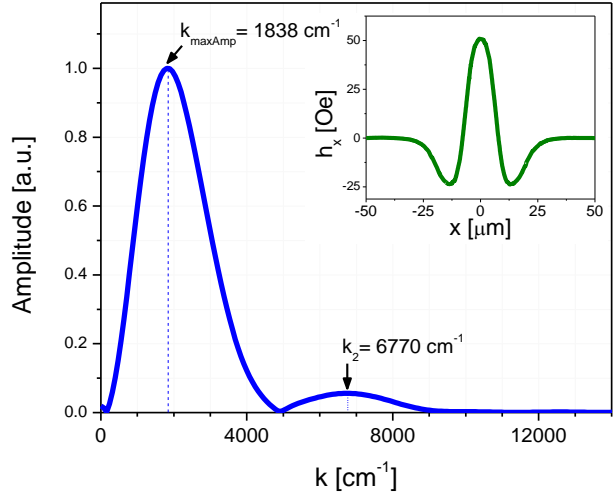

Fig. 6. Excitation spectrum of the CPW with 9.8 $\mu m$ wide signal lines and 4 $\mu m$ gaps. The inset shows in-plane component of the dynamic magnetic field excited by the CPW.

To extend our study, we performed a series of further simulations for the CPW dimensions, which are achievable with electron- and photolithography. We assumed equal widths of signal and ground lines ($W$) as well as equal widths of gaps between them ($G$). The results are presented in Fig. 7. It is found that for the widths $W$ ranging from $300\ nm$ to $40\ \mu m$, the wavenumber $k_{maxAmp}$ vary between $70000\ cm^{-1}$ and $250\ cm^{-1}$, respectively, revealing the CPW wavenumber probing limits. We also note that the gap width significantly affects $k_{maxAmp}$. In order to accurately extrapolate its contribution to $k_{maxAmp}$, we developed empirical formula which incorporates width $G$:

$$k_{maxAmp} = \frac{2.27}{W + 0.6\ G}. \tag{6}$$

The fittings, according to the Eq. 6, are depicted in Fig. 7 with solid lines. We found that Eq. 6 is valid for gap width $0.1\ W < G < 2\ W$. For $G = 0.74\ W$ this formula is equivalent to the one previously proposed in Ref. [ 10 ] ($k_{maxAmp} \approx \pi/2W$).

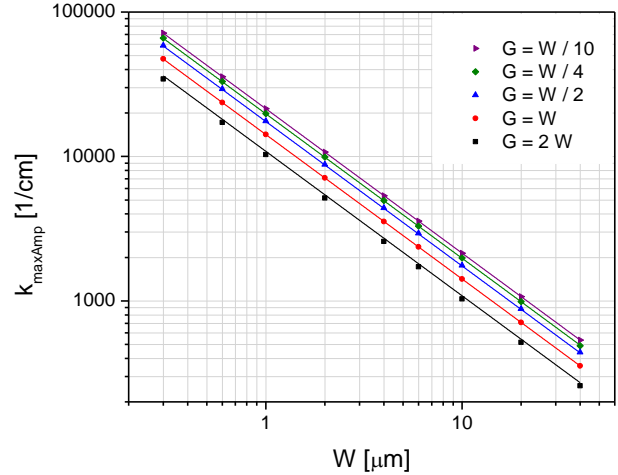

Fig. 7. Wavenumber of the highest amplitude as a function of CPW signal line width. The solid lines represent a fit according to Eq. 6.

To conclude, we reported on long-range spin wave propagation in the 82 $nm$ thick YIG film over the distance as large as 150 $\mu m$. In order to precisely determine excited wavenumber by the coplanar antenna, it is essential to take into account anisotropy fields present in YIG films. We showed that anisotropy significantly affects SW propagation characteristics, namely it causes an increase in SW frequency as well as in SW group velocity. The main contribution comes from the out-of-plane uniaxial anisotropy field. The cubic anisotropy field is negligibly small in the YIG (111) film and it does not affect magnetization dynamics in the film plane. We explained that the wavenumber determination from group velocity vs. magnetic field dependence requires only two types of measurement, that is broadband SW spectroscopy and the measurement of film thickness.

## Acknowledgements

This work was carried out within the Project NANOSPIN PSPB-045/2010 supported by a grant from Switzerland through the Swiss Contribution to the enlarged European Union. J. Rychły and J. Dubowik would like to acknowledge support from the European Union's Horizon 2020 MSCA-RISE-2014: Marie Skłodowska-Curie Research and Innovation Staff Exchange (RISE) Grant Agreement No. 644348 (MagIC). The authors would like to thank Professor Maciej Krawczyk for thoughtful suggestions. We also acknowledge valuable comments from Dr. Piotr Graczyk and Paweł Gruszecki.